\pdfoutput=1
\documentclass[journal]{IEEEtran}
\usepackage[ansinew]{inputenc}
\usepackage{graphicx}
\usepackage{epsfig}
\usepackage{amsmath}    
\usepackage{amssymb}
\usepackage{subcaption}
\usepackage[all]{xy}
\usepackage{multirow}
\usepackage{tikz}
\usetikzlibrary{snakes,matrix,trees,arrows}

\usepackage[a4paper,left=1.8cm,right=1.8cm,top=2.4cm,bottom=2.4cm]{geometry}

\usepackage{float} 

\usepackage{color}
\usepackage{xcolor}

\makeatletter

\def\zapcolorreset{\let\reset@color\relax\ignorespaces}
\def\colorrows#1{\noalign{\aftergroup\zapcolorreset#1}\ignorespaces}

\makeatother

\begin{document}
%

%
\title{All-Path Routing Protocols: \\ Analysis of Scalability and Load Balancing Capabilities for Ethernet Networks\\}

%
%
%

\author{Elisa~Rojas,
        Guillermo~Ibanez,
        Jose~Manuel~Gimenez-Guzman,
        and~Juan~A.~Carral
\thanks{Elisa~Rojas is with Research Department, Telcaria Ideas S.L., 28911, Leganés (Madrid), Spain. e-mail: elisa.rojas@telcaria.com}
\thanks{Guillermo~Ibanez, Jose~Manuel~Gimenez-Guzman and~Juan~A.~Carral are with Departamento de Automática, Edificio Politécnico, University of Alcala, 28871 Alcalá de Henares (Madrid), Spain. e-mails: guillermo.ibanez@uah.es, josem.gimenez@uah.es, juanantonio.carral@uah.es}
}

\markboth{Version 1.00}%
{Rojas \MakeLowercase{\textit{et al.}}: All-Path Routing Protocols: Analysis of Scalability and Load Balancing Capabilities for Ethernet Networks}

\maketitle


\begin{abstract}
This paper presents a scalability and load balancing study of the All-Path protocols, a family of distributed switching protocols based on path exploration.
ARP-Path is the main protocol and it explores every possible path reaching from source to destination by using ARP messages, selecting the lowest latency path. Flow-Path and Bridge-Path are respectively the flow-based and bridge-based versions,
instead of the source address-based approach of ARP-Path. While preserving the main advantages of ARP-Path, Flow-Path has the advantages of full independence of flows for path creation, guaranteeing path symmetry and increased path diversity. While Bridge-Path increases scalability by reducing forwarding table entries at core bridges. We compare the characteristics of each protocol and the convenience of using each one depending on the topology and the type of traffic. Finally, we prove their load balancing capabilities analytically and via simulation.
\end{abstract}

\begin{IEEEkeywords}
Ethernet, Switching, Bridging, Routing bridges, Shortest Path Bridging, Data Centers
\end{IEEEkeywords}

%
\IEEEpeerreviewmaketitle

\section{Introduction}
\label{introduction}

Ethernet switched networks offer the highest performance/cost ratio for local, campus, data center and metro networks, with a high compatibility between elements, and a simpler configuration than IP. Nevertheless, traditional layer 2 protocols either severely limit the network size and performance by blocking redundant links to prevent loops --like the Spanning Tree Protocol (STP/RSTP)~\cite{MACBridges}--, or require additional overhead to compute the paths --like SPB~\cite{Allan12} or TRILL RBridges~\cite{RBridges,Perlman11}--.

Recently, the Software-Defined Networking (SDN) paradigm has unveiled a world of possibilities for Ethernet networks. Popular SDN frameworks, such as OpenDaylight (ODL)~\cite{odl} or Open Network Operating System (ONOS)~\cite{onos,onos-paper}, have developed applications that implement switching protocols. Thanks to their global control of the network components, computing optimal paths is particularly easy. However, SDN still requires to defeat some challenges~\cite{sdn-challenges}, such as scalability issues~\cite{sdn-scalability,sdn-reliability}.

In this situation, simple, distributed, zero configuration protocols that remove the limitations of RSTP and, at the same time, allow scaling Ethernet, might become the key to boost its deployment on campus, data center and enterprise networks. 

The ARP-Path protocol emerged as a shortest path proposal~\cite{IbanezCL} based on the exploration~\cite{Marukawa11} of the network topology without requiring complex link-state protocols, similarly to the ideas shown in~\cite{Cheng04,Shenoy12}. More concretely, ARP-Path is a bridging protocol that finds the lowest latency path to destination. It is based on evolved bridging mechanisms that take advantage of the information conveyed by the ARP protocol message dialog to construct the forwarding table.
ARP-Path is the first protocol of the All-Path family~\cite{Rojas15}. It provides high path diversity, because path selection is sensitive to latency, and it exhibits native load routing with excellent results in throughput~\cite{IbanezCL,Rojas15}. However, some scenarios demand higher scalability or finer path granularity for load balancing, specially when data traffic is highly asymmetric. Flow-Path and Bridge-Path protocols are ARP-Path protocol variants designed to fulfill these requirements. In particular, Flow-Path is able to provide finer load balancing, while Bridge-Path increases the scalability provided by ARP-Path. As all the above-mentioned protocols are based on similar principles, we define the so-called All-Path family, which includes these three protocols.

This paper describes and compares the protocols of the All-Path family in terms of their scalability and load balancing features. In Section~\ref{family} the protocols under study are described in detail, while in Section~\ref{theoretical} we propose an analysis of scalability versus load balancing. In Section~\ref{analytic} we develop an analytical model to evaluate the load distribution in the All-Path family protocols. Afterwards, Section~\ref{related} analyzes the state of the art, as well as possible evolutions of the All-Path family towards a hybrid SDN paradigm. Finally, in Section~\ref{conclusions}, we summarize the main conclusions of the paper.
\section{All-Path Family}
\label{family}

To understand the operation of the All-Path family, we need to describe the ARP-Path protocol first~\cite{IbanezCL}, since it originated the rest of the family and its principles are applicable to the rest of the All-Path protocols.

ARP-Path obtains its name from the Address Resolution Protocol (ARP), invoked in IPv4 prior to any communication between a couple of final hosts, whose messages (ARP Request and ARP Reply) are used to explore the whole network and build a path between those final hosts at the same time. In this way, ARP-Path explores all possible paths in the network and selects the minimum latency path just by snooping the ARP messages, without any change (neither in the messages, nor in the final hosts). Besides, no IP information is needed, therefore the equivalent in IPv6, the Neighbour Discovery Protocol (NDP), could be used in an analogous way to explore those paths.

The operation of ARP-Path is described in the next subsection. The Flow-Path protocol is explained in subsection~\ref{sec:flow}, which follows similar steps in the creation of paths, but it creates unique paths per pairs of hosts or per flow instead of being shared by different final hosts (case of ARP-Path). Finally, as explained in subsection~\ref{sec:bridge}, Bridge-Path generates one path per edge bridge in the topology, which causes groups of hosts connected to a single bridge or switch to share a common path.

\subsection{ARP-Path}
\label{sec:arp}

When a source host $A$ starts a communication with a destination host $B$, $A$ emits an ARP Request message that is replied with an ARP Reply from $B$ containing the MAC address of $B$ previously requested by $A$.

\begin{figure}
        \centering
        \begin{subfigure}[htb]{0.48\textwidth}
                \includegraphics[width=\textwidth]{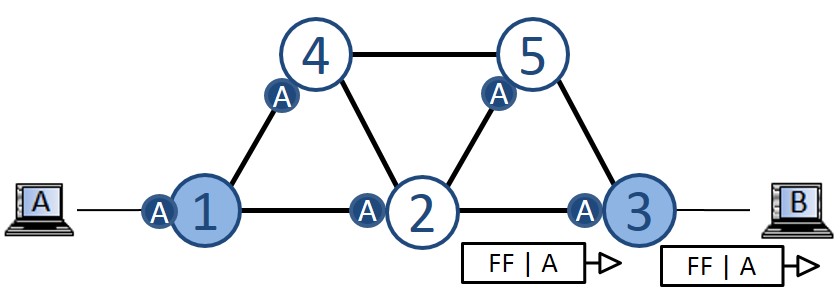}
                \caption{Learning process for path to $A$ by snooping the ARP Request emitted from $A$.}
                \label{ARP_1}
        \end{subfigure}%
        \quad
        \begin{subfigure}[htb]{0.48\textwidth}
                \includegraphics[width=\textwidth]{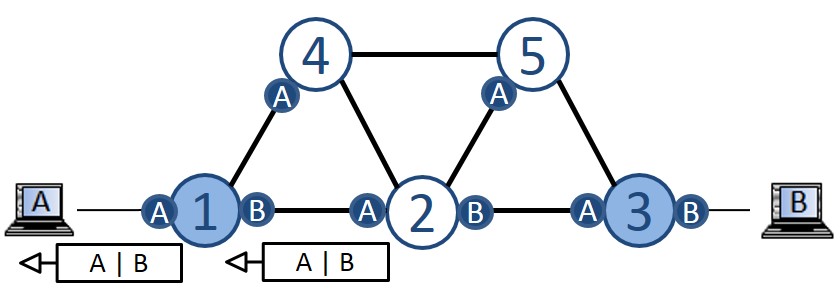}
                \caption{Learning process for path to $B$ by snooping the ARP Reply emitted from $B$ with destination $A$.}
                \label{ARP_2}
        \end{subfigure}
        \begin{subfigure}[htb]{0.48\textwidth}
        \vspace{0.3cm} \small Legend:\\
        	\begin{minipage}[htb]{0.15\linewidth}
						\vspace{0.2cm}
						\includegraphics[width=\textwidth]{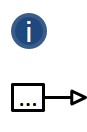}
					\end{minipage}
					\begin{minipage}[htb]{0.85\linewidth}
						\small
						\textit{locked} port, that later will become \textit{learnt}\\(it means a table entry: \textbf{mac}$|$\textbf{port}$|$\textbf{timer}).
						
						\vspace{0.3cm}
						Ethernet frame: \textbf{dst mac} $|$ \textbf{src mac}.

					\end{minipage}
        \end{subfigure}
        \caption{ARP-Path operation.}\label{fig:ARP}
\end{figure}

The first message (ARP Request) has broadcast destination address, so after arriving to the first switch, switch $1$ in Fig.~\ref{ARP_1}, this locks the input port of the frame with the source address of the message, $A$, and sends the frame through all its ports but the one that received it. Then the ARP Request message reaches the switches $2$ and $4$, which carry out the same action, that is, locking the source address to the input port that received the first copy of the frame --i.e. the fastest copy-- and keep broadcasting the frame. When any of the switches receives a later copy of the frame at some other port, this copy is discarded as the path followed by that frame is considered slower. This way, loops are avoided~\cite{Rojas15}.

Finally, one of the message copies, the fastest one, reaches the destination host $B$ after having locked one port in each switch traversed, which means that every switch in the network has the path to $A$, as seen in Fig.~\ref{ARP_1}.

Every locked port is a table entry with four fields: MAC address, associated port, state and timer. After a short time, the entry automatically goes from \textit{locked} to \textit{learnt}. The reason why there are two states is that the first state (\textit{locked}) is needed to avoid loops that might be created by broadcast frames, so it is fixed --no modification allowed-- and has a short timer; while the second (\textit{learnt}) just shows the learnt path to some final host, it is flexible --it is modified based on network changes-- and has a longer timer.

When the destination host $B$ replies to the ARP Request with an ARP Reply, this unicast frame directed to $A$ is able to follow the path to $A$ that has been just explored and, at the same time, to build a path to $B$. To do this, every switch forwards the ARP Reply through the port associated to $A$, as with any other unicast frame, but it also associates the input port to the address of $B$. In this case, the created entry state is directly \textit{learnt} since it is not necessary to prevent loops anymore. Therefore, switch $3$ receives the message, associates $B$ to the input port where the frame was received and sends it through the port associated to $A$, passing through switch $2$ and finally $1$, which operate in the same way, until reaching host $A$, as Fig.~\ref{ARP_2} shows.

After the ARP standard procedure, the communication between $A$ and $B$ starts by means of the previously created path that involves the switches $1\leftrightarrow 2\leftrightarrow 3$. Moreover, those entries (the ones to reach $A$ and the ones to reach $B$) can be shared by third-party hosts, that is, if there was a host $C$ connected to switch $3$, this host could use the same path towards $A$ than the one used by $B$ in Fig.~\ref{ARP_2}, and the same would happen to a host $D$ connected to switch $4$, but in this case the path would be defined by switches $1\leftrightarrow 4$.

\subsection{Flow-Path}
\label{sec:flow}

The Flow-Path protocol subscribes to the same philosophy than ARP-Path: snooping ARP messages to build paths. However, the Flow-Path protocol paths are unique per couple of hosts --or per flow-- and not shared with any other host out of the ARP messages exchange.

\begin{figure}
        \centering
        \begin{subfigure}[htb]{0.48\textwidth}
                \includegraphics[width=\textwidth]{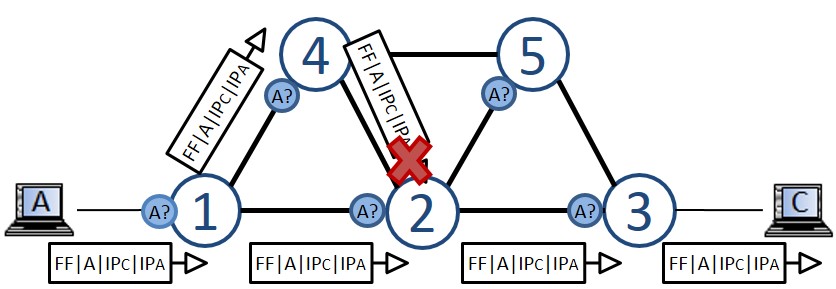}
                \caption{Learning process for path to $A$ of the flow by snooping the ARP Request emitted from $A$.}
                \label{Flow_1}
        \end{subfigure}%
        \quad
        \begin{subfigure}[htb]{0.48\textwidth}
                \includegraphics[width=\textwidth]{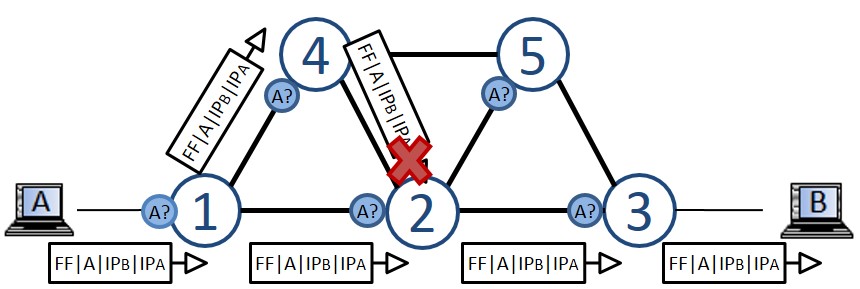}
                \caption{Learning process for path to $B$ of the flow by snooping the ARP Reply emitted from $B$ with destination $A$.}
                \label{Flow_2}
        \end{subfigure}
        \begin{subfigure}[htb]{0.48\textwidth}
        \vspace{0.4cm} \small Legend:\\
        	\begin{minipage}[htb]{0.15\linewidth}
						\vspace{0.2cm}
						\includegraphics[width=\textwidth]{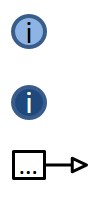}
					\end{minipage}
					\begin{minipage}[htb]{0.85\linewidth}
						\small
						\textit{locked} port, (it means a table entry: \textbf{mac\_src}$|$\textbf{mac\_dst}$|$\textbf{ip\_src}$|$\textbf{ip\_dst}$|$\textbf{port}$|$\textbf{timer}).
						
						\vspace{0.2cm}
						
						\textit{confirmed} port, (it means a table entry: \textbf{mac\_src}$|$\textbf{mac\_dst}$|$\textbf{ip\_src}$|$\textbf{ip\_dst}$|$\textbf{port}$|$\textbf{timer}).
						
						\vspace{0.3cm}
						Ethernet frame: \textbf{dst mac} $|$ \textbf{src mac} $|$ \textbf{dst ip} $|$ \textbf{src ip}.

					\end{minipage}
        \end{subfigure}
        \caption{Flow-Path operation.}\label{fig:Flow}
\end{figure}

Figure~\ref{Flow_1} shows how switches lock the ports belonging to the path between $A$ and $B$ directed to $A$. Since $B$'s MAC address is still unknown\footnote{In fact ARP aims to discover $B$'s MAC address}, Flow-Path temporarily writes down the IP addresses of hosts $A$ and $B$ in order to distinguish the flow from any other in which $A$ also participates. Meanwhile, the entry is shown as $A?$ where the question mark refers to $B$'s address, which will be known after receiving the corresponding ARP Reply.

As observed in Fig.~\ref{Flow_2}, in an analogous way to ARP-Path, the ARP Reply message makes switches learn the ports of the path between $A$ and $B$ directed to $B$, denominated $BA$ and, at the same time, confirms those named $A?$ changing their state from \textit{locked} to \textit{learnt} and their value to $AB$ as the destination MAC address is not unknown anymore. Once the path is set, communication between $A$ and $B$ can start and the path is defined by switches $1\leftrightarrow 2\leftrightarrow 3$.

The difference with ARP-Path is that if another host $C$ connected to switch $3$ wanted to send traffic to $A$, the path from $C$ to $A$ might not be the same (as the one from $B$ to $A$) as now they are created independently, that is, the path between $C$ and $A$ would create entries named $AC$ and $CA$ and those could be coincident with the ports of $AB$ and $BA$ or not, depending on the already existing traffic in the network, since the minimum latency path might be a different one.

Thus, Flow-Path guarantees the independence of flows which, at the same time, can guarantee a better distribution of the load in the network in case that certain host exchanges messages with more than one destination. However, the disadvantage of this proposal is that forwarding tables are bigger and independent paths might not be required if traffic is low.

\subsection{Bridge-Path}
\label{sec:bridge}

The Bridge-Path protocol is based on the opposite idea to Flow-Path: instead of creating independent paths per flow to balance the load, the objective is to share the paths even with more hosts than with ARP-Path by building routes per edge switch (which is connected to a group of hosts) and not per individual final hosts. In this way, forwarding tables are smaller, which guarantees higher scalability.

There are three variants in order to deploy this protocol without having to modify the ARP messages:
\begin{itemize}
\item Reusing the VLAN tag (ARP-PathV).
\item Encapsulating the frame with MAC-in-MAC (ARP-PathM).
\item Translating the host address into a hierarchical address in which certain part or field has the ID of the edge switch (Path-Moose~\cite{IbanezIEICE}).
\end{itemize}

The first and second variants follow the same basics of encapsulation than SPBV and SPBM respectively~\cite{Allan12}, while the third one is based on the MOOSE protocol~\cite{Scott10}.

To explain the operation of Bridge-Path, we consider the specific case of ARP-PathM, but note that any other variant would be analogous. When a host $A$ wants to communicate with a host $B$, the message emitted by the source (being it an ARP or not) is encapsulated in the switch that serves it with a new Ethernet header, which indicates source and destination with a MAC address field that formats some type of ID of the edge switch. This encapsulated frame enters the network and the rest of the switches will operate in the same way that with ARP-Path (they do not necessarily know that the frame is encapsulated with MAC-in-MAC and that the MAC addresses represent IDs of the edge switches instead of hosts), until the frame reaches the switch serving the destination host, which decapsulates it and sends the original frame to the destination host. That is, the only difference resides on edge switches, which are required to encapsulate and decapsulate in order to generate grouped paths.

\begin{figure}
        \centering
        \begin{subfigure}[htb]{0.48\textwidth}
                \includegraphics[width=\textwidth]{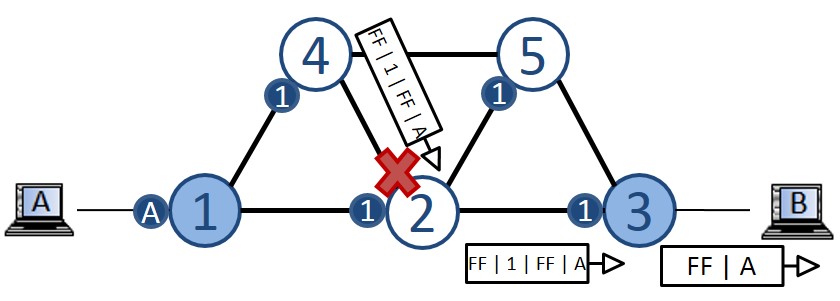}
                \caption{Learning process for path to bridge $1$ (edge of $A$) by snooping the ARP Request emitted from $A$.}
                \label{Bridge_1}
        \end{subfigure}%
        \quad
        \begin{subfigure}[htb]{0.48\textwidth}
                \includegraphics[width=\textwidth]{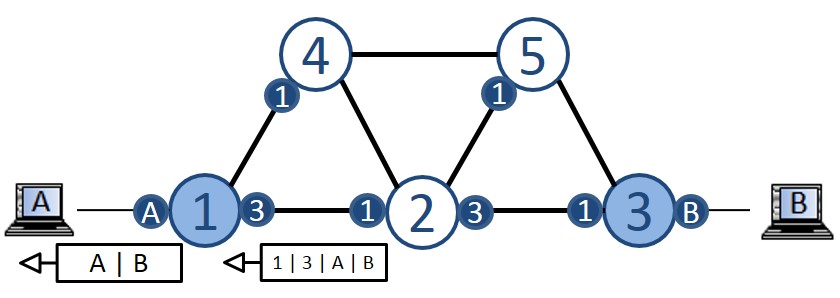}
                \caption{Learning process for path to bridge $3$ (edge of $B$) by snooping the ARP Reply emitted from $B$ with destination $A$.}
                \label{Bridge_2}
        \end{subfigure}
        \begin{subfigure}[htb]{0.48\textwidth}
        \vspace{0.3cm} \small Legend:\\
        	\begin{minipage}[htb]{0.15\linewidth}
						\vspace{0.2cm}
						\includegraphics[width=\textwidth]{figs/ARP_Path3.jpg}
					\end{minipage}
					\begin{minipage}[htb]{0.85\linewidth}
						\small
						
						\textit{locked} port, that later will become \textit{learnt}\\(it means a table entry: \textbf{mac/id}$|$\textbf{port}$|$\textbf{timer}).
						
						\vspace{0.3cm}
						Ethernet frame: \textbf{dst mac} $|$ \textbf{src mac}.

					\end{minipage}
        \end{subfigure}
        \caption{Bridge-Path operation.}\label{fig:Bridge}
\end{figure}

Bridge-Path's path learning operation is shown in Figs.~\ref{Bridge_1} and~\ref{Bridge_2}, respectively. Broadcast messages do not change the destination MAC address after encapsulation (remaining FF:FF:FF:FF:FF:FF, $F\!F$ for short in the figure), but they do change the source (from $A$ to $1$, which might be the MAC of the edge switch or an ID of it, in the figure). In the case of unicast messages, both addresses are translated into their corresponding edge switches. In Fig.~\ref{Bridge_2} the address of host $B$ is translated into $3$ and the address of host $A$ into $1$, which is known thanks to the previous ARP Request, and forwarding is done based on $3$ and $1$, thus ignoring the encapsulated addresses $B$ and $A$, respectively. Note that edge switches need to save the information about other edge switches and their connected hosts in order to proceed. This information is conveyed by the ARP messages.

In Bridge-Path, if there were a host $C$ connected to switch $3$ interested in communicating with host $A$, it would share the same path from $B$ to $A$, which is the one indicated by the entries of address $1$, similarly to ARP-Path. The difference though is that if there were a host $D$ connected to the edge bridge $1$, the path from $C$ to $D$ would still be the same as the one from $B$ to $A$, which is not necessarily true in ARP-Path but it will always happen in Bridge-Path. As in the Bridge-Path protocol routes are shared by several hosts, scalability is improved with respect to ARP-Path, in exchange for worse load balancing capabilities.
\section{Comparison of the All-Path Family Protocols}
\label{theoretical}

After describing the different All-Path protocols in the previous section, a few conclusions can be easily drawn. Flow-Path is expected to achieve better load balancing since it is able to create more than a single path per final host, followed by ARP-Path with one path per host and finally Bridge-Path, which on average builds less than one path per host because routes are shared among the set of hosts attached to a common switch. However, the size of the forwarding tables is also much bigger for Flow-Path, followed by ARP-Path and the smallest for Bridge-Path, being this a crucial parameter in order to evaluate scalability. Obviously, table sizes are proportional to the number of paths created per host. Consequently, in this section we analyze the suitability of each protocol of the All-Path family for different topologies to reach a good tradeoff between both capabilities: load balancing and scalability (in terms of table sizes).

\subsection{Load Balancing Analysis}

In this section, we take as a reference of load balancing the theoretical total number of independent paths that a protocol can build per host, in order to compare the three All-Path protocols. The reason for using this parameter to measure the load balancing capabilities is explained in detail in the next section, where we analitically prove that All-Path protocols use all possible paths evenly (since All-Path protocols choose the lowest latency paths and this type of creation tends to select the less used resources).
The number of independent bidirectional paths that can be created by Flow-Path, ARP-Path and Bridge-Path on average, denominated $P_{FP}$, $P_{AP}$ and $P_{BP}$ respectively, is:


\begin{equation}
P_{FP}=F_B=\frac{H \cdot (H-1)}{2}
\label{PFP}
\end{equation}

\begin{equation}
P_{AP}=\frac{H}{2}
\label{PAP}
\end{equation}

\begin{equation}
P_{BP}=\frac{B_E}{2}
\label{PBP}
\end{equation}

Being:
\begin{itemize}
\item $F_B$: average number of bidirectional flows in the network.
\item $H$: average number of active hosts.
\item $B_E$: average number of active edge switches ($B_E \leq H$, since an active edge switch is always attached to one or more active hosts).
\end{itemize}

Note that we are considering bidirectional paths, that is, the resources used in both directions of the communication. In a communication between a host $A$ and a host $B$, even if the path from $A$ to $B$ is different than the path from $B$ to $A$, in practice we can consider it a bidirectional path in terms of resources being used, thus simplifying the analysis.

As the equations show, the number of independent paths that Flow-Path can create is the highest, followed by ARP-Path and finally Bridge-Path ($P_{FP} \geq P_{AP} \geq P_{BP}$), as expected. In order to measure the load balancing capability, we will compare this theoretical value with the actual number of available paths in the networks, since the number of theoretical paths is bounded by the actual number of possible paths in the network ($\Psi$).

\subsection{Scalability Analysis}

For this analysis, we will refer to the total number of table entries required in all the switches of the network, as it is the only difference among the All-Path protocols regarding this parameter.

Thus, the total number of table entries in the network created on average by Flow-Path, ARP-Path and Bridge-Path, denominated $T_{FP}$, $T_{AP}$ and $T_{BP}$ respectively, is:

\begin{equation}
T_{FP}=F_U \cdot b=H \cdot (H-1) \cdot b
\label{TFP}
\end{equation}

\begin{equation}
T_{AP}=H \cdot (b+L_e)
\label{TAP}
\end{equation}

\begin{equation}
T_{BP}=B_E \cdot (b+L_e)
\label{TBP}
\end{equation}
Being:
\begin{itemize}
\item $F_U$: average number of unidirectional flows in the network ($F_U = 2 \cdot F_B$, since a bidirectional flow can be seen as two unidirectional flows and each direction of the flow needs a table entry).
\item $H$: average number of active hosts.
\item $B_E$: average number of active edge switches ($B_E \leq H$, since an active edge switch is always attached to one or more active hosts).
\item $b$: average number of switches that form a path for a flow or couple of hosts.
\item $L_e$: average number of switches that also share the path to the same destination from different sources (note that $L_e$ switches are not included in $b$).
\end{itemize}

Note that we have chosen to represent these last equations as a function of the average number of unidirectional flows, instead of the bidirectional flows, because they are more easily to deduce in this way, but it is possible to substitute $F_U = 2 \cdot F_B$ if we want them to depend on the same parameter.

In this case, Flow-Path generates a higher number of table entries than ARP-Path, and ARP-Path create more entries than Bridge-Path; being the results proportional to the square of $H$
for Flow-Path, to $H$ for ARP-Path and, finally, to a fraction of $H$ for Bridge-Path.

Another parameter to take into account is the average number of switches that form the path, $b$, since the three equations are a function of it. However, ARP-Path and Bridge-Path in fact depend on the addition of $b$ and $L_e$ because paths are shared, i.e. when a flow creates a path (defined by $b$ switches on average), different sources can join the already existing paths just by adding branches (defined by $L_e$ switches on average), defining a tree in the end ($b+L_e$), while Flow-Path generates a single path per each established communication.

If we calculate the quotient between the previous equations, we obtain the following ratios:

\begin{equation}
R_{FA}=\frac{T_{FP}}{T_{AP}}=\frac{H \cdot (H-1) \cdot b}{H \cdot (b+L_e)}= (H-1)\cdot \frac{b}{b+L_e}
\label{RFA}
\end{equation}

\begin{equation}
R_{AB}=\frac{T_{AP}}{T_{BP}}=\frac{H \cdot (b+L_e)}{B_E \cdot (b+L_e)}= \frac{H}{B_E} \geq 1
\label{RAB}
\end{equation}

As shown in Eq.~\ref{RFA}, the ratio $R_{FA}$ between the number of table entries of Flow-Path and ARP-Path does not only depend on the average active hosts $(H-1)$,
but it also depends on the network shape ($\frac{b}{b+L_e}$): the wider the network is, the higher will be $L_e$ and the lower the ratio $R_{FA}$. While Eq.~\ref{RAB} shows that the relationship $R_{AB}$ between ARP-Path and Bridge-Path will always be greater than or equal to $1$, depending on the average number of hosts per edge switch, as we expected.

\subsection{Numerical Evaluation}
\label{analysis}

With the objective of assessing which is the best protocol to be used in network routing, we have evaluated the three above-mentioned All-Path protocols in two different meshed network topologies.

\begin{figure}[h]
\centering
\includegraphics[width=0.2\textwidth]{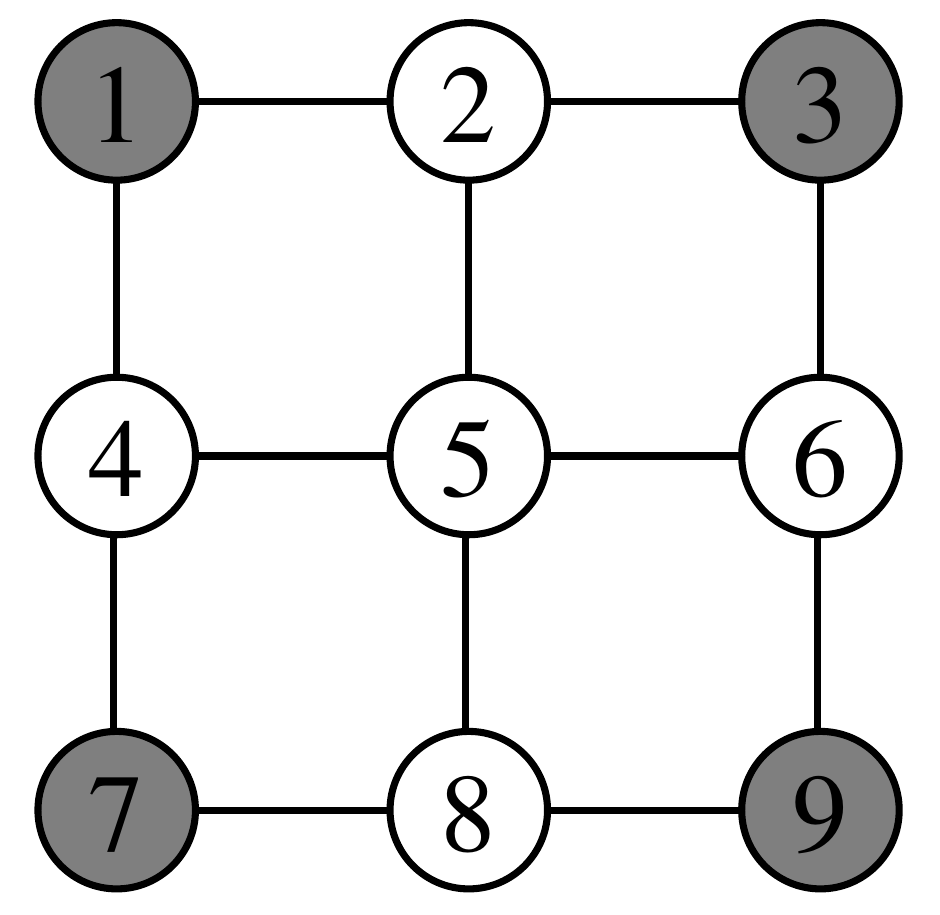}
\caption{Simple grid topology $n \times n$, with $n=3$.}
\label{fig:simple_grid}
\end{figure}

\subsubsection{Simple grid network topology}
\label{simple_grid}

The first network topology under study is a simple grid with size $n \times n$. In Fig.~\ref{fig:simple_grid} we show an example of that topology for $n=3$, i.e. a topology with $4$ edge bridges. Note that in that figure shaded nodes represent edge bridges, i.e. those bridges connected to other bridges and final systems, while white nodes represent core bridges, i.e. those bridges that are connected only to other bridges. We will study the number of table entries and the number of paths as a function of the topology size $n$. At the same time, $n$ affects parameters $b$ and $L_e$. We also consider $H$, the mean number of active hosts in the topology.

\begin{figure*}
        \centering
        \begin{subfigure}[h]{0.53\textwidth}
                \includegraphics[width=\textwidth]{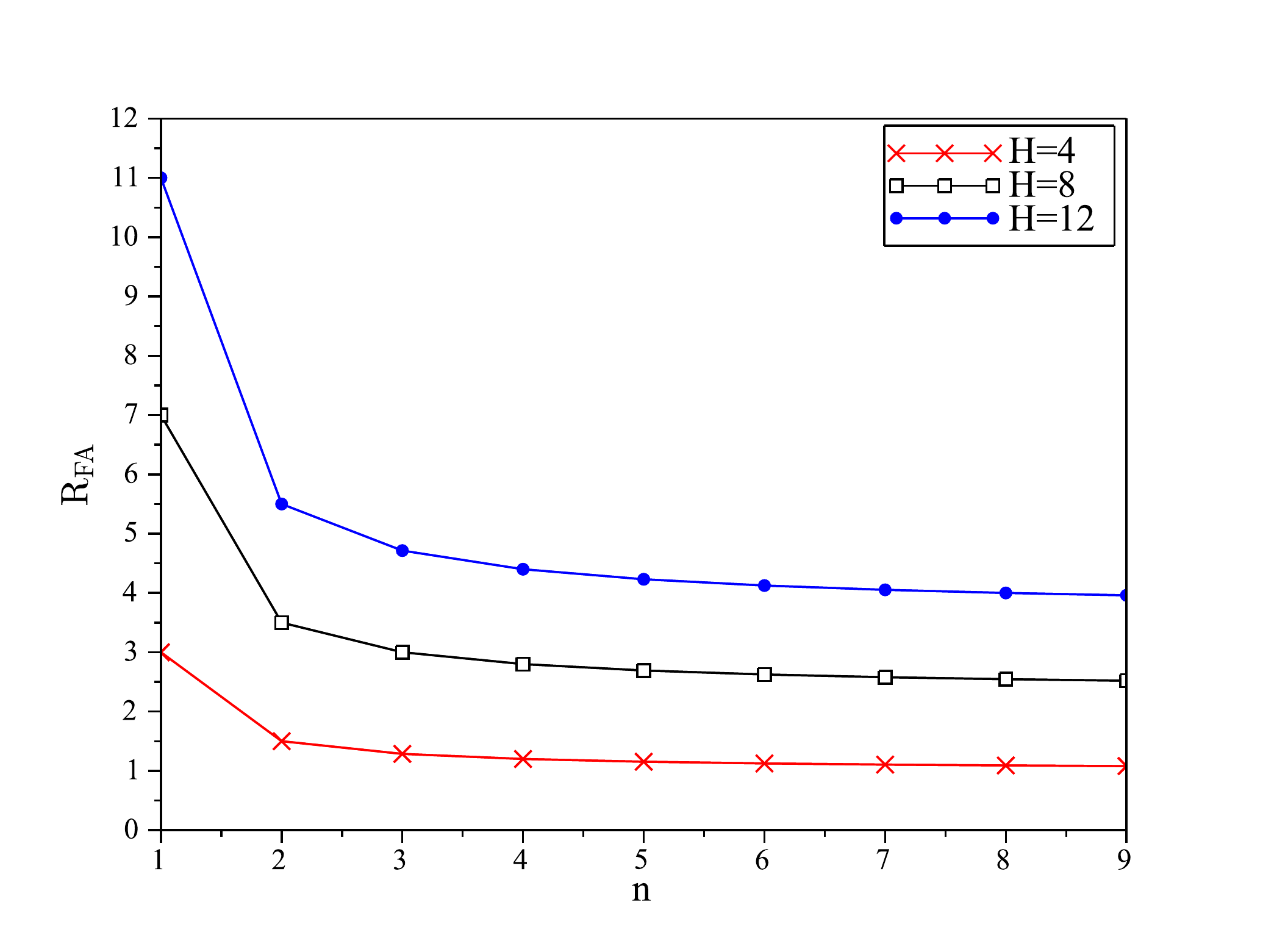}
                \caption{$R_{FA}=T_{FP}/T_{AP}.$}
                \label{TFP_TAP}
        \end{subfigure}%
        \begin{subfigure}[h]{0.53\textwidth}
                \includegraphics[width=\textwidth]{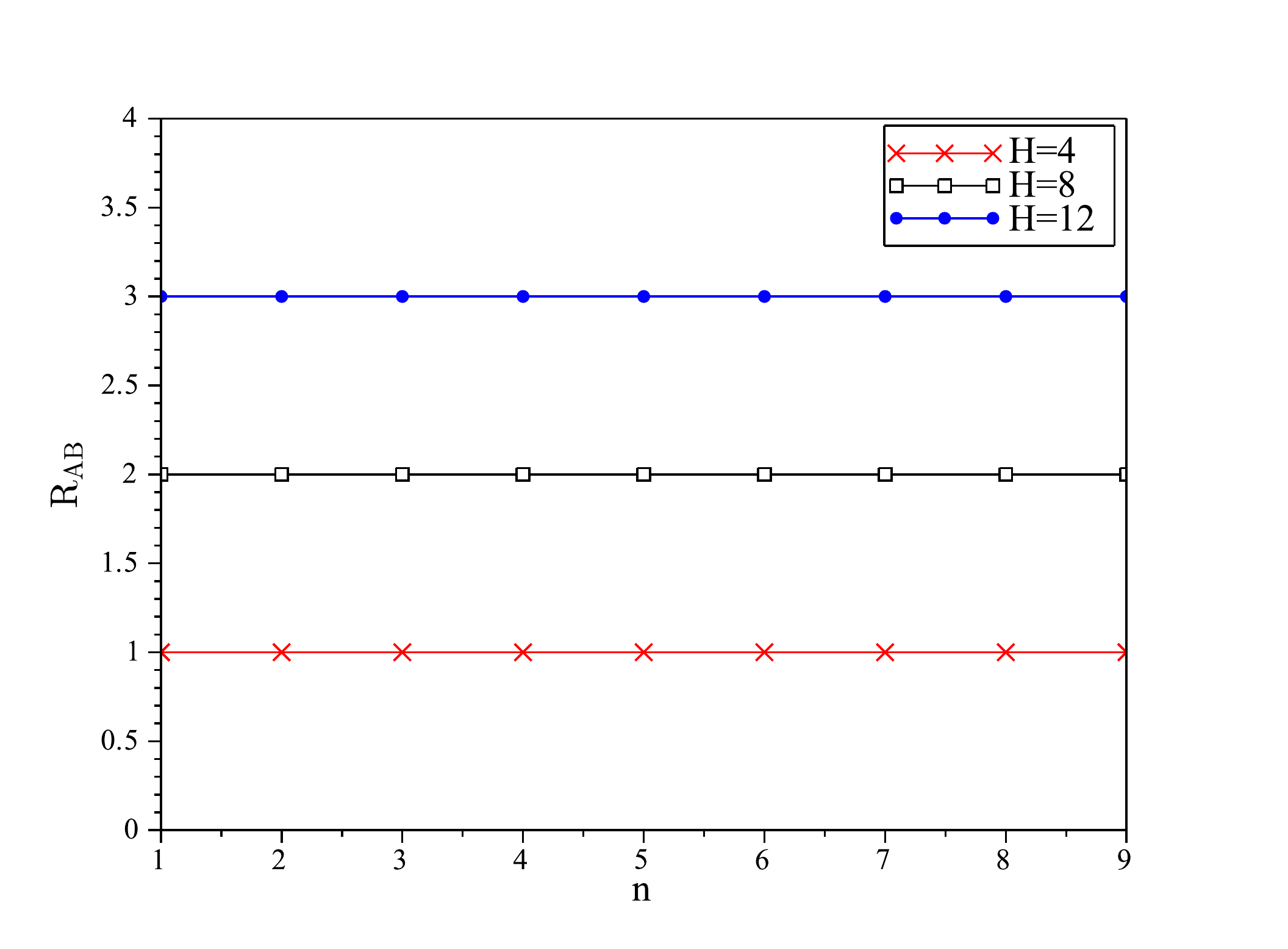}
                \caption{$R_{AB}=T_{AP}/T_{BP}.$}
                \label{TAP_TBP}
        \end{subfigure}
        \caption{Ratio between the number of table entries as a function of $n$ (size) and $H$ (average active hosts in the topology).}\label{fig:tablesizes}
\end{figure*}

As it can be seen from Fig.~\ref{fig:tablesizes}, the ratio between the number of table entries between Flow-Path and ARP-Path decreases as the topology size increases and becomes wider. For example, when $H=12$ (three active hosts on average in every one of the four edge routers), that ratio is different from $11$ (i.e. $H-1$) as one could intuitively think in advance. Instead, the ratio is closer to $4$ as the topology ($n$) increases. This is because, as stated in Eq.~\ref{RFA}, the network shape (which is the factor $\frac{b}{b+L_e}$) also affects $R_{FA}$. Meanwhile, the ratio between ARP-Path and Bridge-Path is $1$, $2$ and $3$ for respectively $H=4$, $H=8$ and $H=12$, which is the average number of hosts per edge bridge.

To explore the possible paths, we have taken into account the paths between the opposite sides of the network, i.e. between bridges $1$ and $9$ or $3$ and $7$ (Fig.~\ref{fig:simple_grid}). In the case of possible paths we have considered only the shortest paths. For example, for $n=2$ there are $2$ shortest paths (from $3$ bridges), for $n=3$ there are $6$ paths (from $5$ bridges), for $n=4$ there are $20$ possible shortest paths, and so on, being this increase exponential. 

\begin{figure}
        \centering
        \begin{subfigure}[tb]{0.53\textwidth}
                \includegraphics[width=\textwidth]{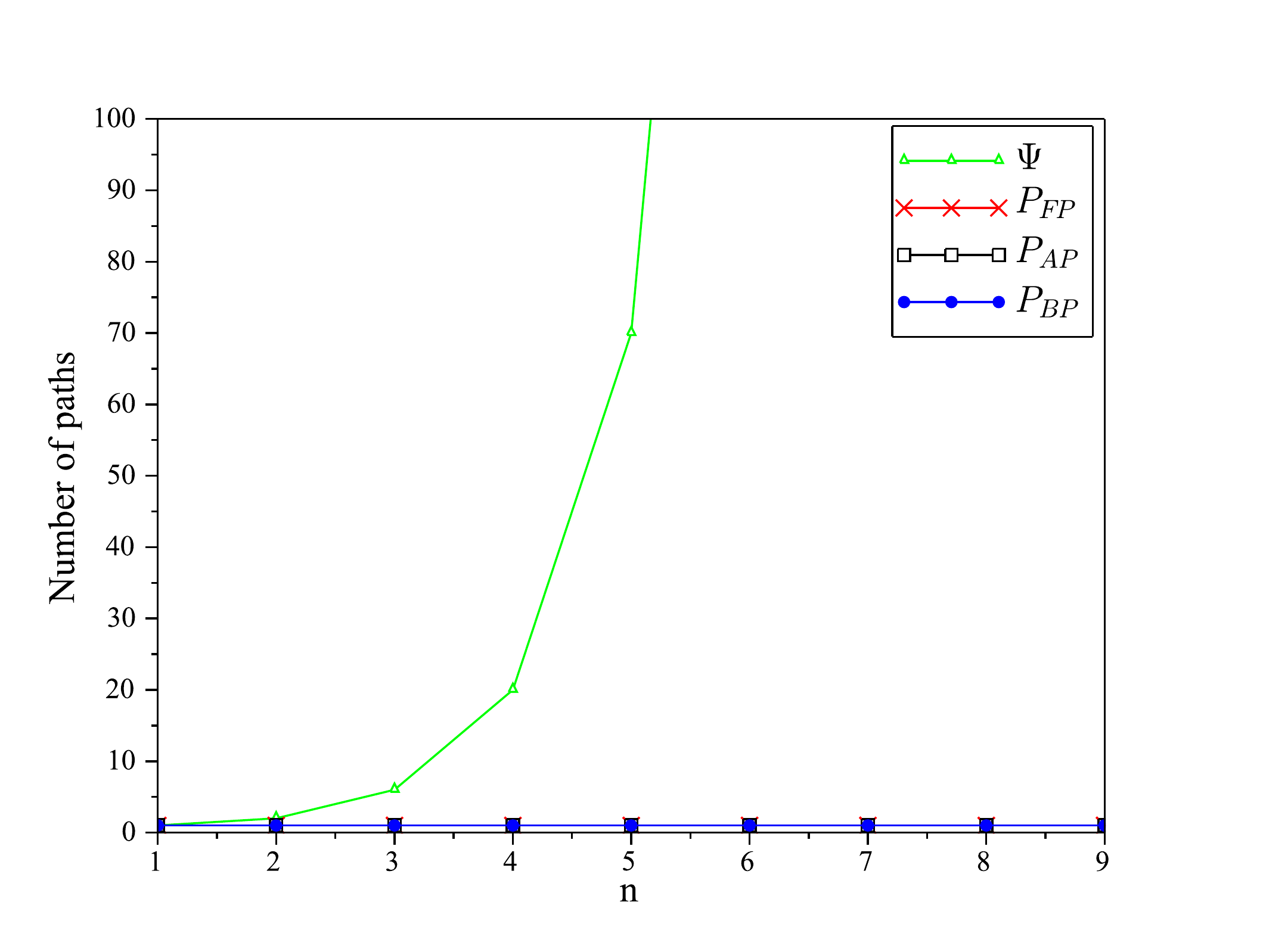}
                \caption{$H=4.$}
                \label{normal_H4}
        \end{subfigure}\\%
        \begin{subfigure}[tb]{0.53\textwidth}
                \includegraphics[width=\textwidth]{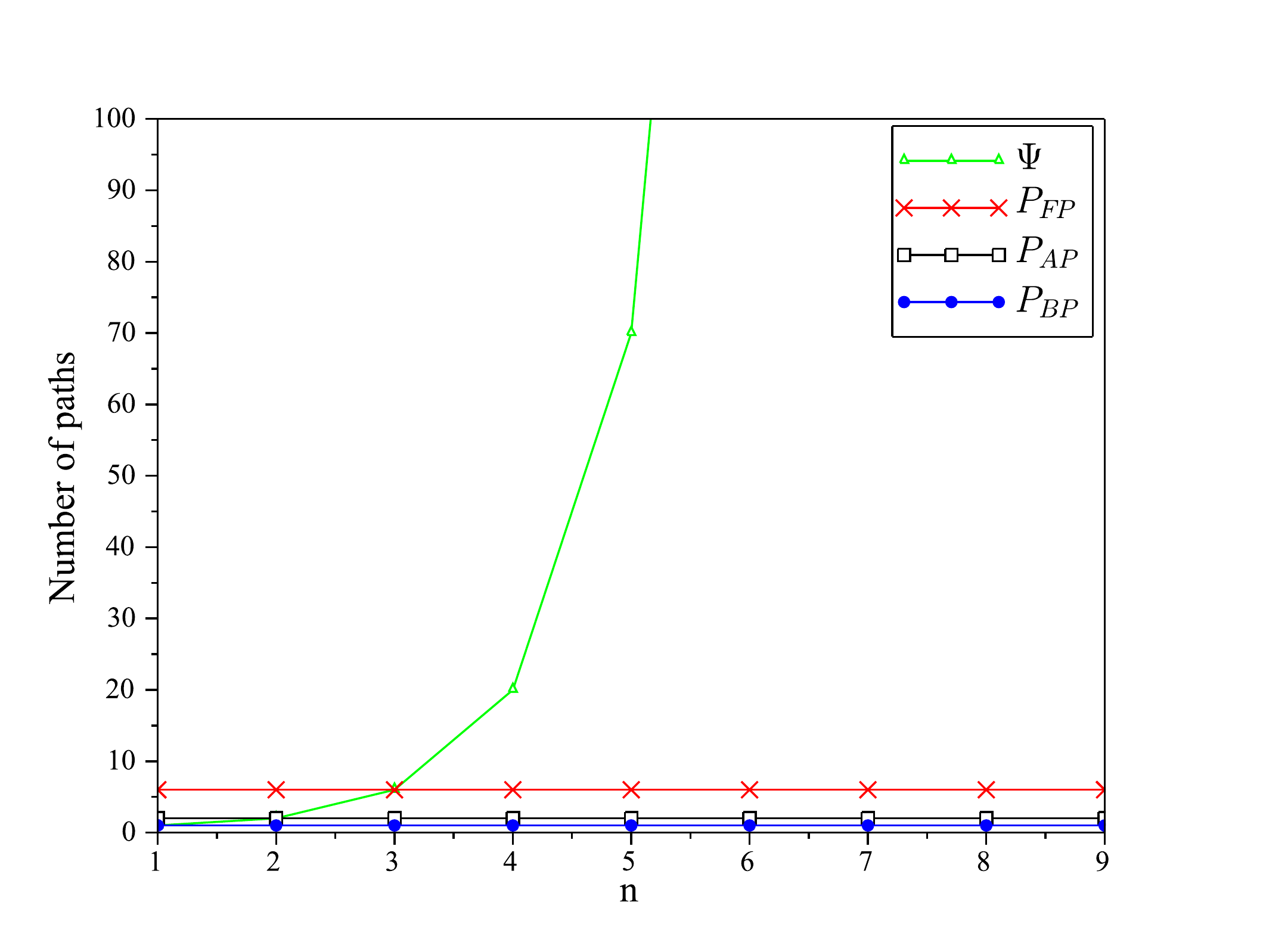}
                \caption{$H=8.$}
                \label{normal_H8}
        \end{subfigure}\\%
        \begin{subfigure}[tb]{0.53\textwidth}
                \includegraphics[width=\textwidth]{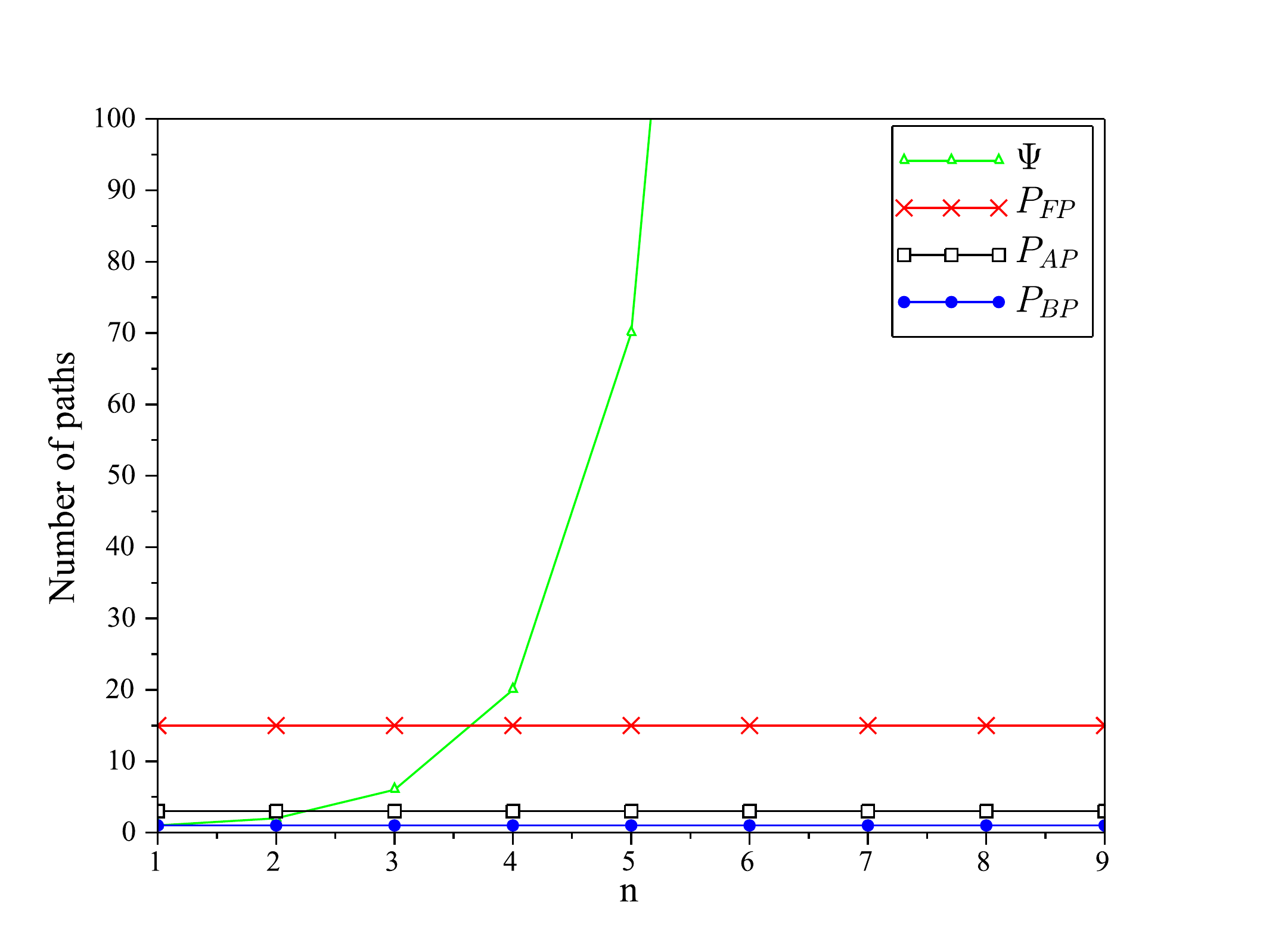}
                \caption{$H=12.$}
                \label{normal_H12}
        \end{subfigure}%
        \caption{Available paths and paths created by each protocol in simple grid network topology.}\label{normal}
\end{figure}

From Fig.~\ref{normal}, we can conclude that, as $H$ increases, Flow-Path becomes more suitable, mainly as $n$ is higher, because path diversity increases about $10$ times in relation to the one of ARP-Path or Bridge-Path with table sizes only $4$ times larger. However, for smaller values of $H$ and $n$, the best choice is Bridge-Path, with a much lower cost.

\begin{figure}[h]
\centering
\includegraphics[width=0.2\textwidth]{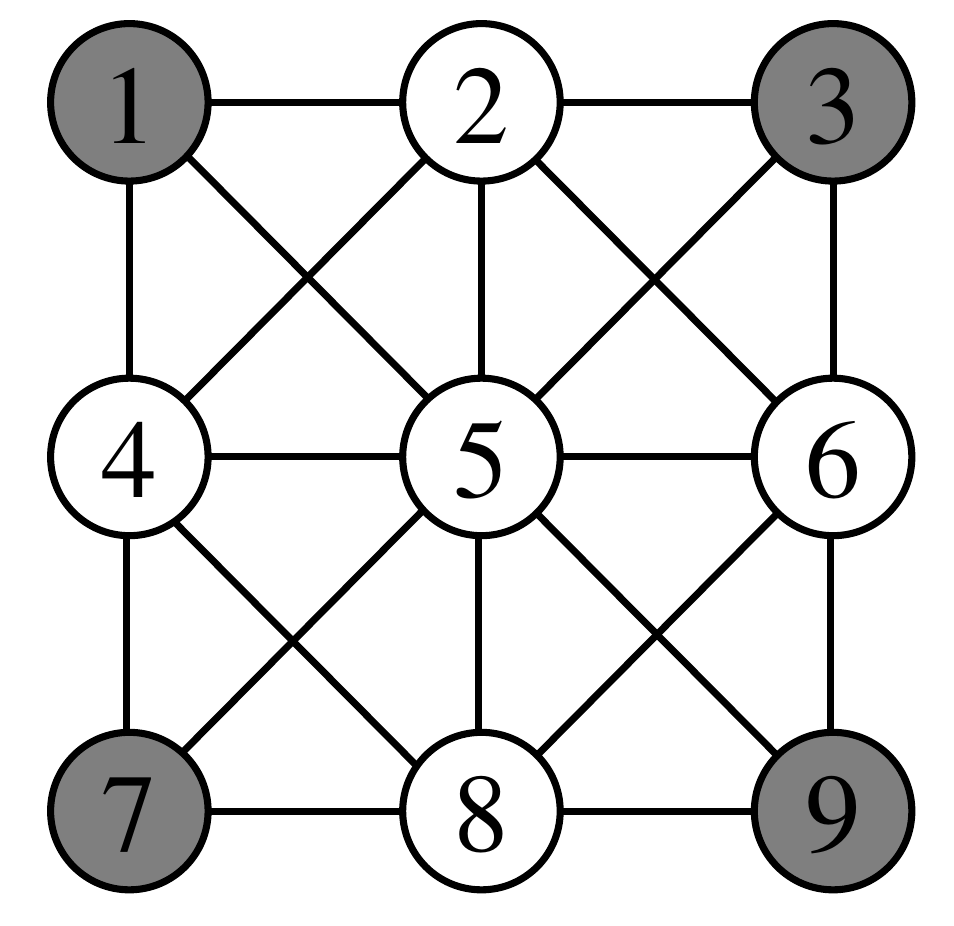}
\caption{Crossed grid topology $n \times n$, with $n=3$.}
\label{fig:crossed_grid}
\end{figure}

\subsubsection{Crossed grid network topology}
\label{crossed_grid}

\begin{figure}
        \centering
        \begin{subfigure}[htb]{0.53\textwidth}
                \includegraphics[width=\textwidth]{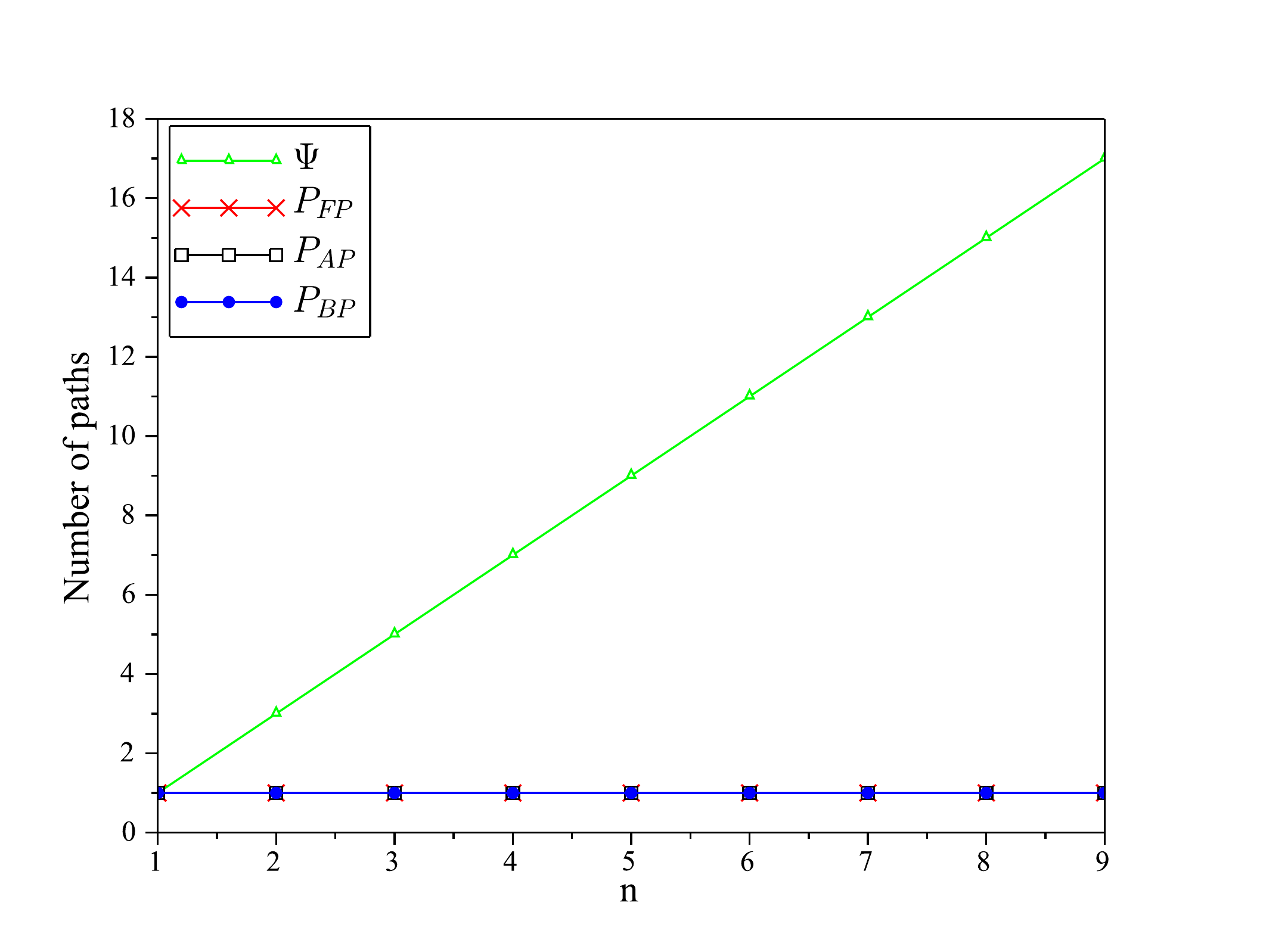}
                \caption{$H=4.$}
                \label{cruzada_H4}
        \end{subfigure}%
        \quad
        \begin{subfigure}[htb]{0.53\textwidth}
                \includegraphics[width=\textwidth]{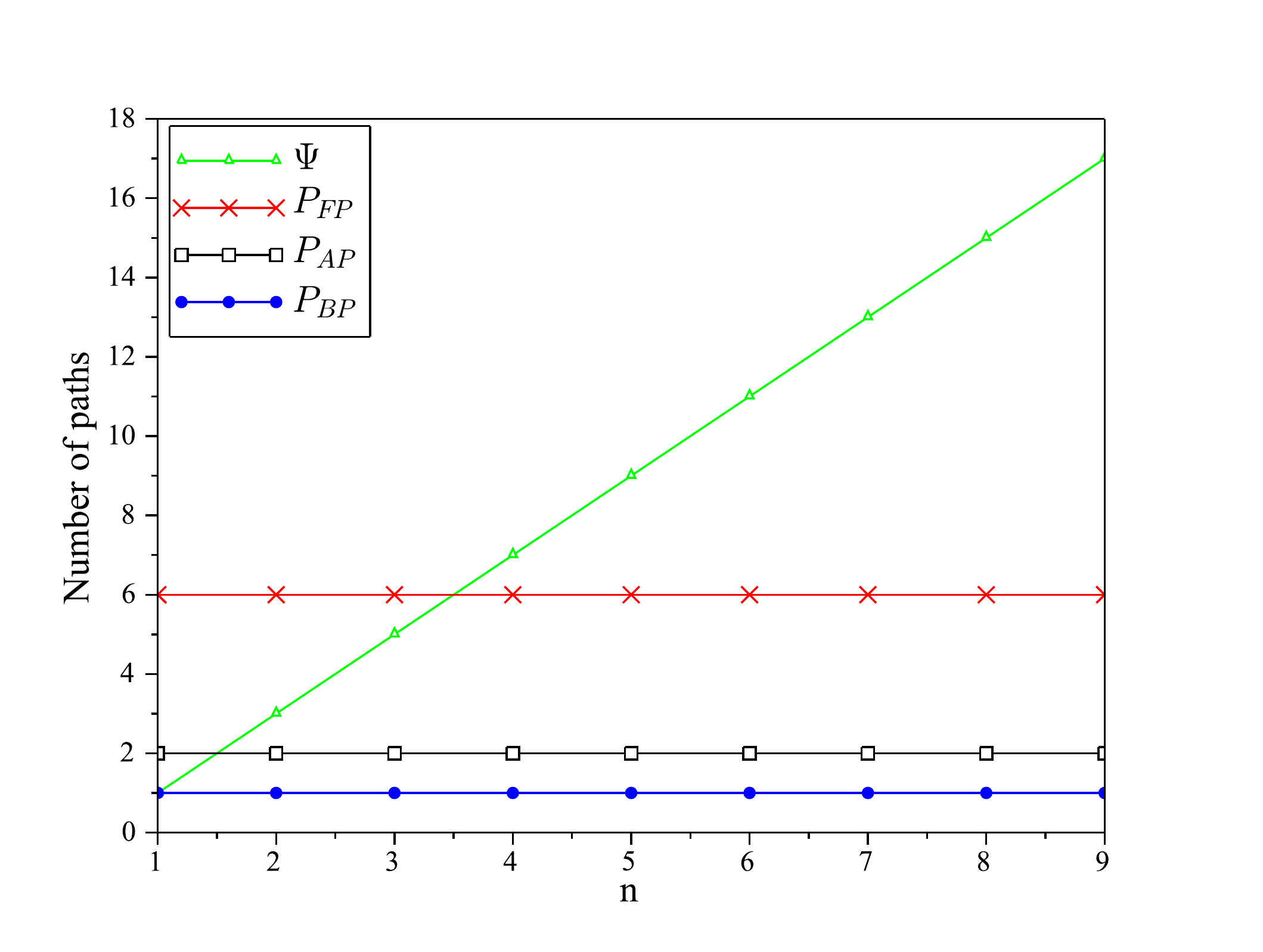}
                \caption{$H=8.$}
                \label{cruzada_H8}
        \end{subfigure}%
        \quad
        \begin{subfigure}[htb]{0.53\textwidth}
                \includegraphics[width=\textwidth]{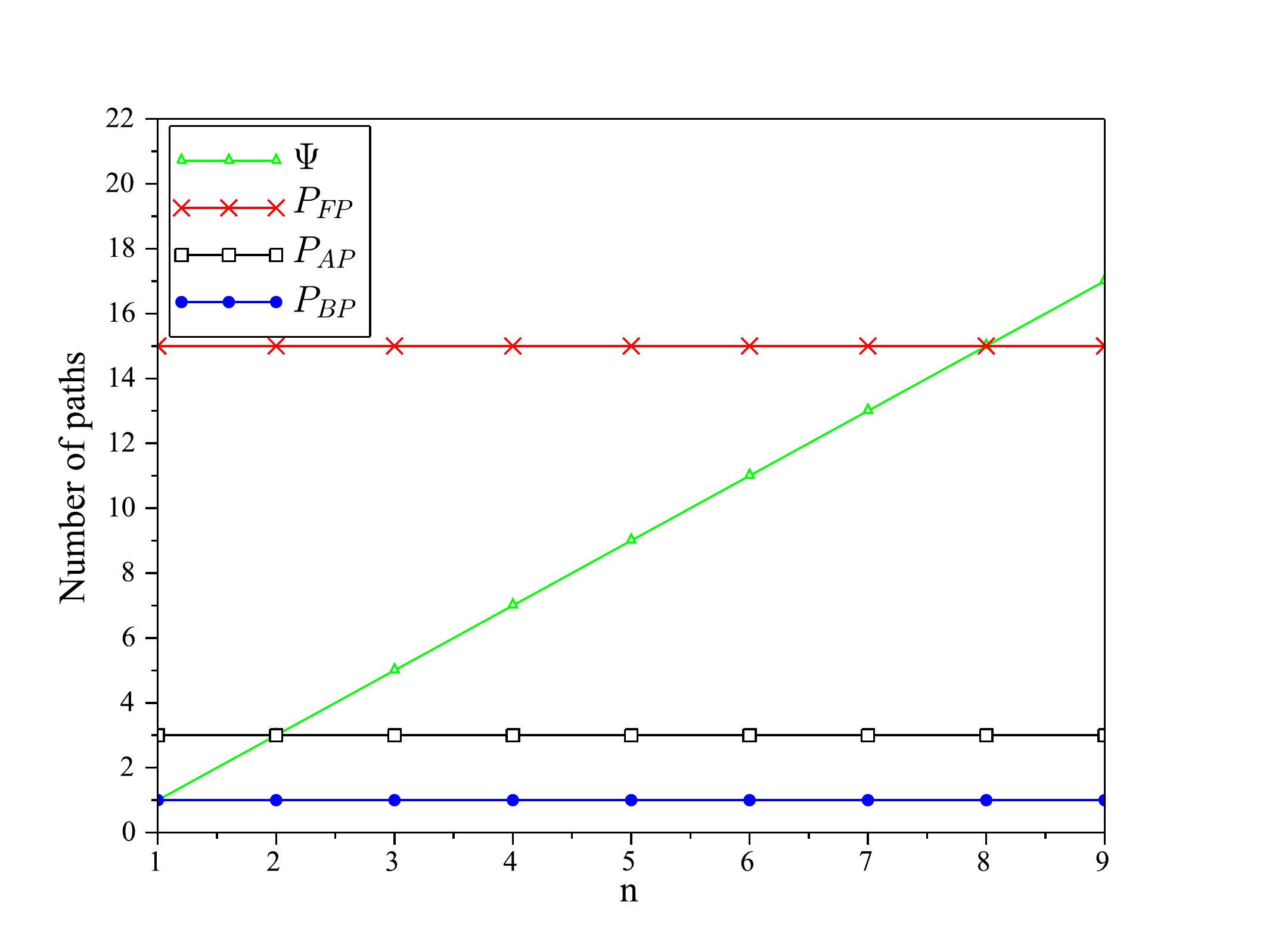}
                \caption{$H=12.$}
                \label{cruzada_H12}
        \end{subfigure}%
        \caption{Available paths and paths created by each protocol in crossed grid network topology.}\label{cruzada}
\end{figure}

Now we consider a topology that is similar to the previous one, but including crossed diagonal links between bridges, as shown, for $n=3$, in Fig.~\ref{fig:crossed_grid}. As in Fig.~\ref{fig:simple_grid}, shaded nodes represent edge bridges while white nodes represent core bridges.
The main peculiarity of this topology in comparison with the simple grid is that now there only exists a shortest path, which is the one that traverses the main central diagonal of the grid from one end to the other. Notwithstanding, the ratio between the number of table entries remains the same (as shown in Fig.~\ref{fig:tablesizes}). 

If we consider as possible paths the shortest one and also all those paths that have one more hop than the shortest one (we exclude longest paths, as it is unlikely to use them, although not impossible if the rest are heavily loaded), we obtain the results shown in Fig.~\ref{cruzada}. This figure shows that there are some cases where Flow-Path is not necessary, as the number of generated paths is higher than the number of possible paths in the topology, so we can save table entries and properly share by just using ARP-Path, for example.
\section{Load Distribution Analysis in the All-Path Family}
\label{analytic}

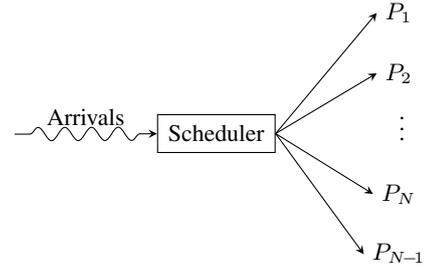
\begin{figure}[tb]
\centering
\small
\tikzstyle{state}=[draw,circle,inner sep=1pt]
\tikzstyle{rect}=[draw,rectangle,inner sep=4pt]
\tikzstyle{transition}=[->,>=stealth]	
\begin{tikzpicture}
\matrix [ampersand replacement=\&,
					         column sep={6mm,between origins},  
					         row sep={8mm,between origins}]
					{
					\& \& \& \& \node(sep1){$P_1$};   \\
					\& \& \& \& \node(sep2){$P_2$};   \\
					\node(sep)[rect]{Scheduler};\\
					\& \& \& \& \node(sep3){$P_N $};  \\
					\& \& \& \& \node(sep4){$P_{N\!-\!1}$};   \\
					};
\draw[->,>=stealth,snake=snake,line after snake=2mm,line before snake=2mm] (-105pt,0pt) -- (sep) node [above,midway] {Arrivals};
\draw[->,>=stealth] (sep.east) -- (sep1.west);
\draw[->,>=stealth] (sep.east) -- (sep2.west);
\draw[->,>=stealth] (sep.east) -- (sep3.west);
\draw[->,>=stealth] (sep.east) -- (sep4.west);
\draw (40pt,5pt) node(x) {$\vdots$};
\end{tikzpicture}
\normalsize
\caption{System model.}
\label{fig:model}
     
\end{figure}

In the previous section, we have used the number of possible paths as the parameter to measure the load balancing capabilities of the different All-Path protocols. In this section, we analytically show how the procedure followed to build a path under an All-Path protocol results in an even load distribution across a network, i.e. when there are several paths with similar features all of them are equally used.
For this purpose, the way that an All-Path protocol sets up a path can be modeled as follows. As shown in Fig.~\ref{fig:model}, new flows, which arrive to the system at mean rate $\lambda$ and request a holding time with rate $\mu$, are routed to any of the possible paths $P_i$, being $N$ the number of possible paths $(P_1,...,P_N)$ between source and destination. We define $L_i$ as the capacity of link $i$, $l_i(t)$ as the available capacity of link $i$ at time $t$ and $C_i$ as the maximum capacity of a path ($C_i$), which is determined by its lowest capacity link, as it acts as a bottleneck,
\[
C_i = min(L_j), \forall j \in P_i.
\]

The scheduling policy of any All-Path protocol is based on the selection of the path with the lowest latency.
The latency of a path can be computed as the sum of the latencies of all links of a path. Note that a link can belong to several paths simultaneously.
For each hop in the path, the latency that a packet will experience is the sum of the transmission, propagation, queueing and processing delays ($d_{trans}$, $d_{prop}$, $d_{proc}$ and $d_{queue}$, respectively). 
We can postulate that both $d_{prop}$ and $d_{proc}$ are independent of the system load, so we can omit them in our analysis. However, the sum of $d_{queue}$ and $d_{trans}$ will highly depend on load. Basically, choosing the lowest latency path is equivalent to choosing the path with the highest number of resources available, because as the available throughput increases, $d_{trans}$ decreases and queues are shorter, so $d_{queue}$ also decreases. For that reason, we have assumed in our analysis that All-Path protocols choose the path with maximum available capacity, which is expected to have the minimum delay (which constitutes the real operation of the protocols).

Given the above description, the behavior of the system can be described by a discrete-state continuous-time process. We can represent the state of the system at any given time by a vector
\[
\mathcal{S}:=\{s_1,s_2,\ldots,s_N\},
\]
where $s_i$ is the available capacity of path $i$, $s_i \in [0,\ldots,C_i]$, which is determined by the capacity of its most congested link, $s_i(t)=min(l_j(t)), \forall j \in P_i$.

To model the scheduling policy of All-Path protocols, we introduce a scheduling policy that is a mixture between a deterministic and a random policy that can be explained as follows:
\begin{enumerate}
\item An arriving flow is always sent to the path with the highest available capacity, i.e. $P_i$ so that $max(s_i(t))$.

\item If the maximum capacity is not unique, the scheduler selects the path randomly among the paths with the maximum available capacity.

\end{enumerate}

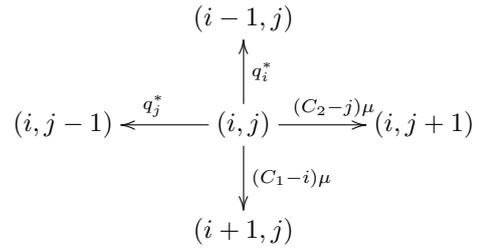
\begin{figure}[tb]
\[
\xymatrix{
& (i-1,j) & \\
(i,j-1) & 
(i,j) \ar[d]^{(C_1-i)\mu} \ar[r]^{(C_2-j)\mu} \ar[u]_{q_i^*} \ar[l]_{q_j^*} & 
(i,j+1) \\
& (i+1,j) & \\
}
\]
\caption{Transition diagram.}
\label{fig:trans}
\end{figure}

For the sake of mathematical tractability we consider the number of paths to be $N=2$. Although this choice is a simplified scenario, it is worth noting that it represents the essence of the path setup in All-Path protocols. We also make the common assumptions of exponentially distributed random variables for the inter-arrival and holding times of the flows with parameters $\lambda$ and $\mu$, respectively.
However, we have also studied more realistic distributions for the parameters that describe the arrival and holding times by means of simulation, as they are not analitically tractable.
Under the abovementioned assumptions, we can represent the state of the system at any given time by a vector $\mathcal{S}:=\{s_1,s_2\}:0\leq s_1\leq C_1; 0\leq s_2\leq C_2$, where $s_i$ is the available capacity of path $i$ ($0<s_i<C_i$). Without loss of generality, we consider that each flow occupies one resource unit, so $C_i$ in this section is measured in resource units. This system is therefore a Continuous Time Markov Chain (CTMC) whose transitions rates are described in Fig.~\ref{fig:trans}, being
\[
q_i^*=
\begin{cases}
\lambda ,& i > j\\
\lambda/2 ,& i = j\\
0 ,& i < j\\
\end{cases} 
\qquad \qquad
\textrm{and}
\qquad \qquad
q_j^*=
\begin{cases}
\lambda ,& j > i\\
\lambda/2 ,& i = j\\
0 ,& j < i\\
\end{cases}
\]

This system constitutes a level dependent Quasi Birth and Death process (QBD)~\cite{neuts81} whose infinitesimal generator ($\mathbf{Q}$) has a block tridiagonal structure with $(C_1+1)\times(C_1+1)$ blocks with size $(C_2+1)\times(C_2+1)$ each:

\begin{align*}
\mathbf{Q}=
\left[ \begin{array}{ccc c ccc  ccc}
\mathbf{D}_0 & \mathbf{M}_0 & \mathbf{0} & \mathbf{0} & \mathbf{0} & \hdots  & \mathbf{0} & \mathbf{0} & \mathbf{0} \\
\mathbf{L}_1 & \mathbf{D}_1 & \mathbf{M}_1 & \mathbf{0} & \mathbf{0} & \hdots & \mathbf{0} & \mathbf{0} & \mathbf{0}  \\
\mathbf{0} & \mathbf{L}_2 & \mathbf{D}_2 & \mathbf{M}_2 & \mathbf{0} & \hdots & \mathbf{0} & \mathbf{0} & \mathbf{0}  \\
\mathbf{0} & \mathbf{0} & \ddots & \ddots  &  \ddots   & \hdots & \hdots & \hdots & \hdots\\
\mathbf{0} & \mathbf{0} & \mathbf{0} & \mathbf{0} & \mathbf{0} & \hdots & \mathbf{L}_{C_1} & \mathbf{D}_{C_1} & \mathbf{M}_{C_1} \\
\end{array} \right]
\end{align*}

The stationary probability distribution can be obtained by solving $\pi\mathbf{Q}=\mathbf{0}$ along with the normalization condition. As $\mathbf{Q}$ is a finite matrix, this system can be solved by any of the standard methods defined in classical linear algebra. However, we can exploit the block tridiagonal structure of $\mathbf{Q}$ using the algorithm 0 defined in \cite{servi02}, which allows us to reduce the computational cost, although there are other proposals useful for that purpose like \cite{gaver84,folding94}.
 
\begin{figure}[hbt]
\centering
\includegraphics[width=0.5\textwidth]{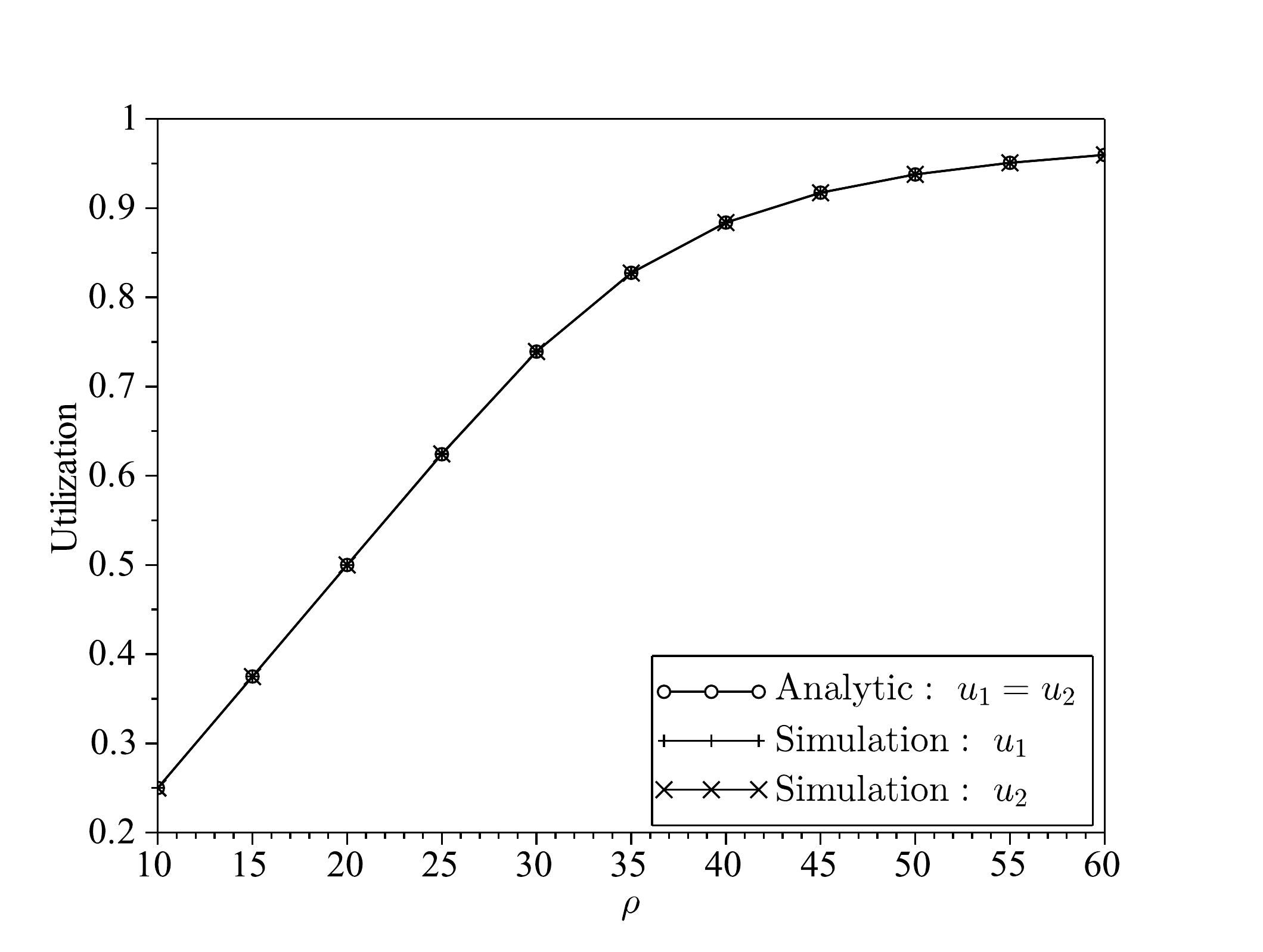}
\caption{Validation of the analytic model with simulation.}
\label{fig:analytic_simul2}
\end{figure}

\begin{figure*}
        \centering
        \begin{subfigure}[h]{0.53\textwidth}
                \includegraphics[width=\textwidth]{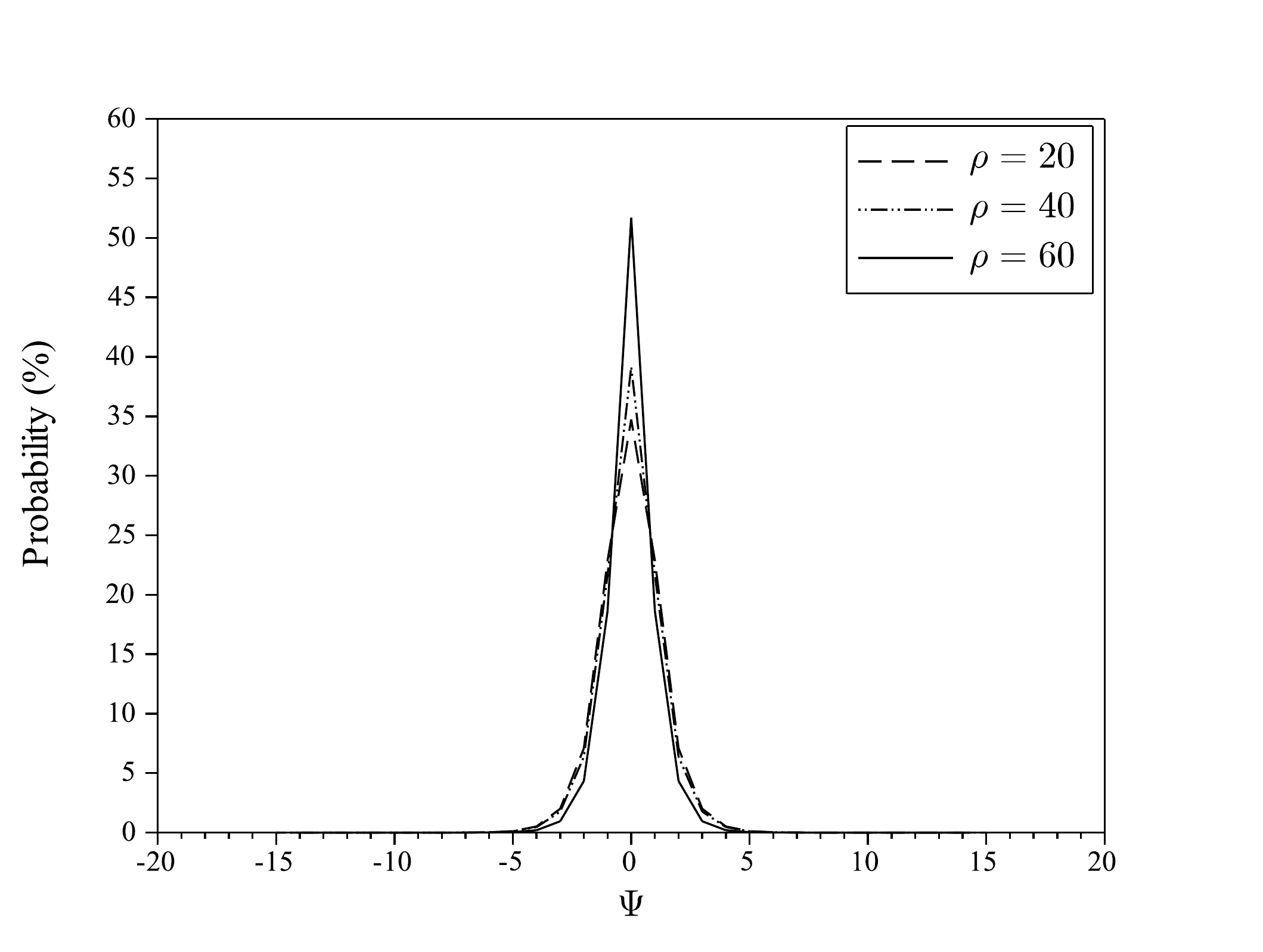}
                \caption{$C_1=C_2=20$.}
                \label{psi_20_20}
        \end{subfigure}%
        \begin{subfigure}[h]{0.53\textwidth}
                \includegraphics[width=\textwidth]{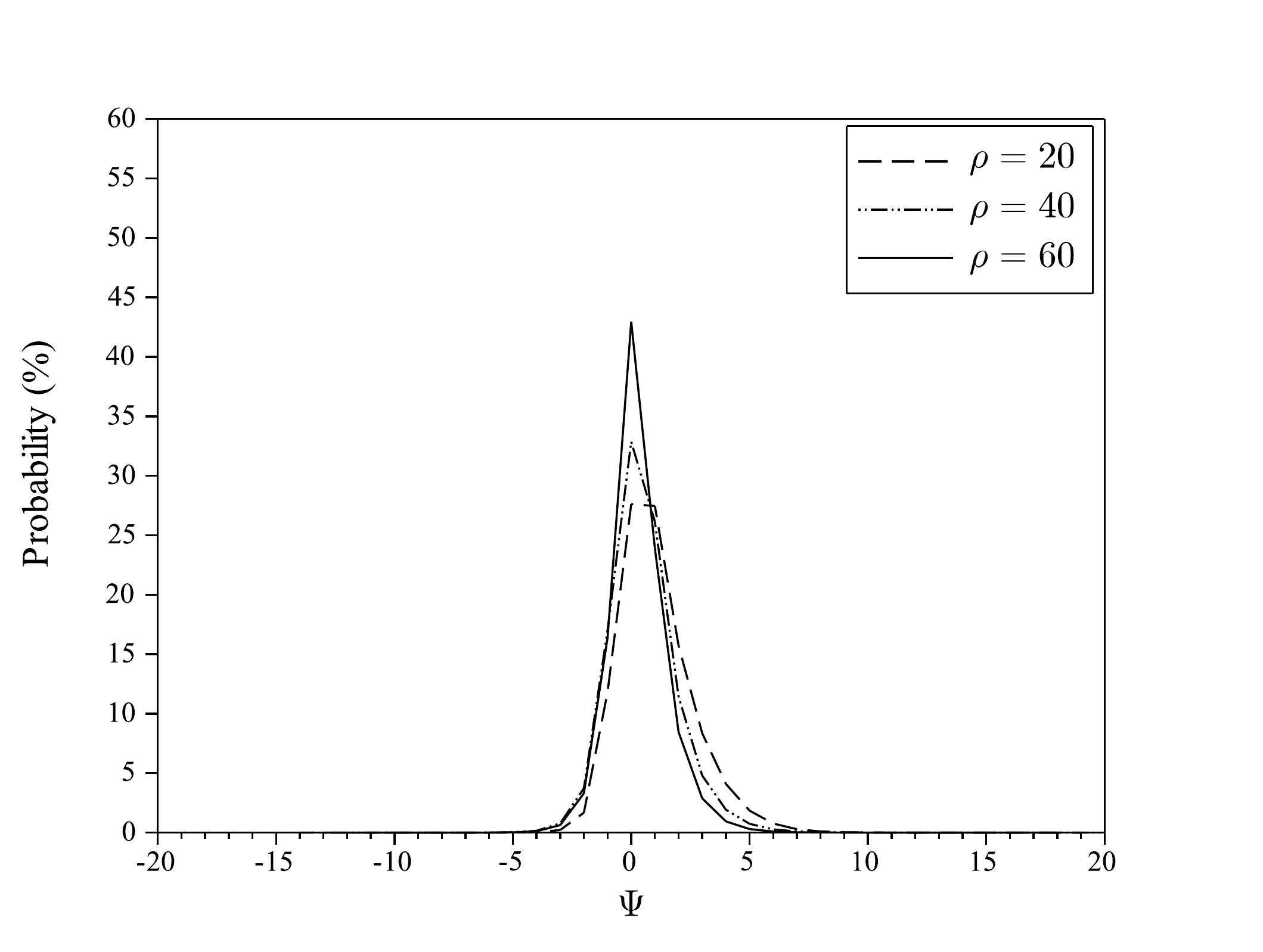}
                \caption{$C_1=30$, $C_2=20$.}
                \label{psi_30_20}
        \end{subfigure}
        \caption{Probability of having a different available capacity (of $\Psi$ resource units) between paths.}\label{fig:results_analytic}
\end{figure*}

In Figs.~\ref{fig:analytic_simul2} and~\ref{fig:results_analytic} we show the main results obtained solving this model for $\mu=1$ and for different values of the offered load to the system $\rho=\lambda/\mu$, so the system operates in very different working points. Figure~\ref{fig:analytic_simul2} validates the analytic model by means of a simulated model, where we have chosen $C_i=20,\forall i$. This figure shows that the utilization ($u_i$) in both models (analytic and simulated) for the two paths (note that in the analytical model the utilization of paths $1$ and $2$ are the same, i.e. $u_1=u_2$) coincide. Once validated the analytical model, we can go in depth of the problem of load distribution studying the probability of having a different available capacity of $\Psi$ resource units between the different paths. In other words, in Fig.~\ref{fig:results_analytic} we show for $\Psi=0$ the probability of both paths having the same available capacity, and for $\Psi=i$ ($-i$) we represent the probability that path $1$ ($2$) has $i$ more available resource units than path $2$ ($1$). Figure~\ref{psi_20_20} stands for $C_1=C_2=20$, whereas Fig.~\ref{psi_30_20} shows the results for a scenario with $C_1=30$ and $C_2=20$.
As it can be concluded from both figures, the probability of being in a state where there is a path with much more resources than the other (high values of $|\Psi|$) is negligible, so the load is properly distributed in order to get the highest available bandwidth (minimum delay).

\begin{figure}[hbt]
\centering
\includegraphics[width=0.5\textwidth]{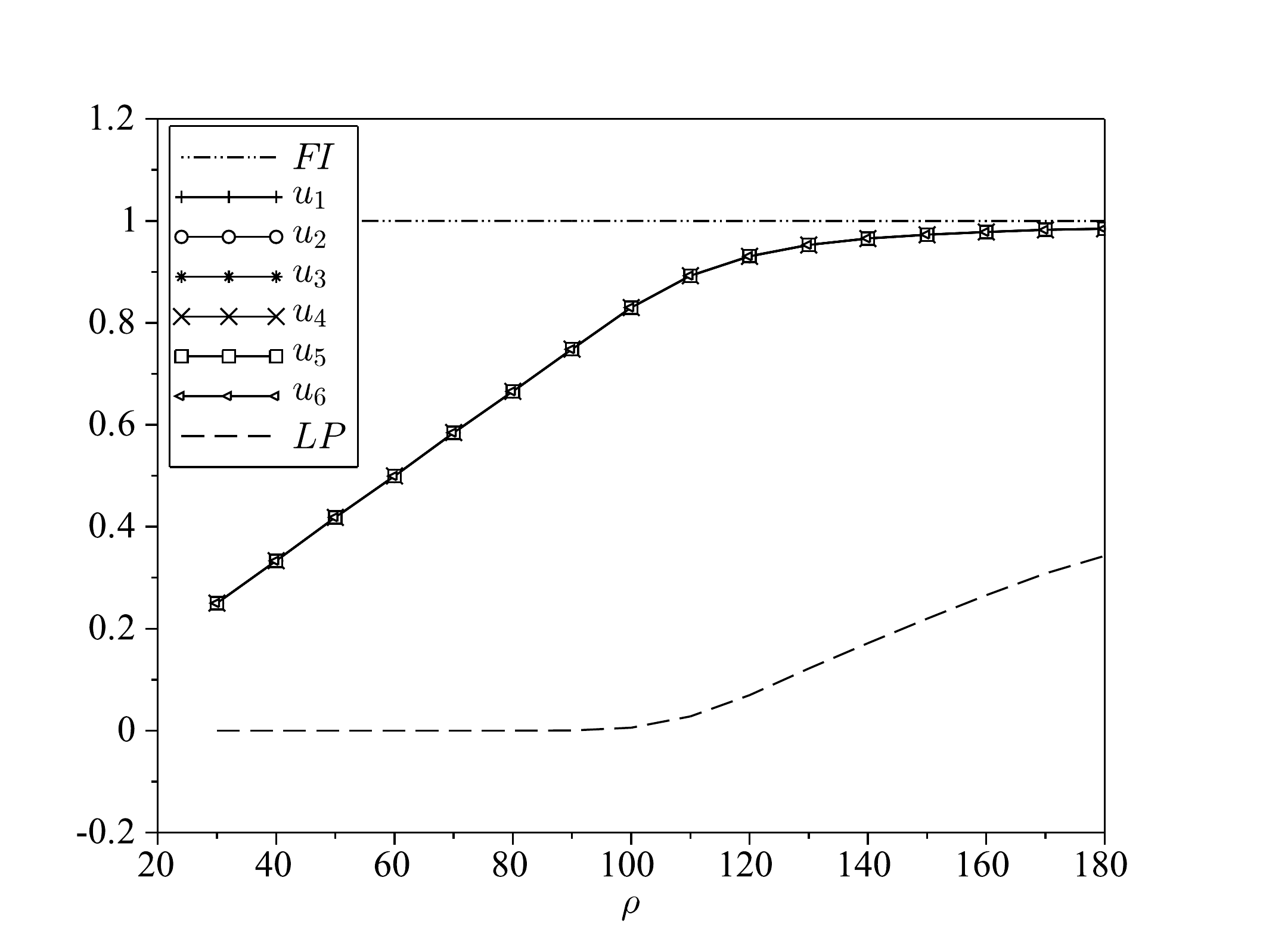}
\caption{Load distribution with $N=6$ and realistic data center traffic.}
\label{fig:simul6}
\end{figure}

In order to evaluate if the All-Path protocols are able to attain an even load distribution in more realistic scenarios, we have simulated a more complex scenario with $N=6$ possible paths. This situation could represent a simple grid scenario with size $3\times 3$ as the one shown in Fig.~\ref{fig:simple_grid} or a state-of-the-art data center topology, where there are $6$ shortest paths. For the underlying traffic that is transported by the network we have considered realistic data center traffic. In this type of networks, and similarly to Internet flow characteristics, there are myriads small flows (usually called mice) and a small number of large flows (elephants), transporting these last ones most of the traffic~\cite{VL2}. From~\cite{VL2}, we have considered in our simulator that only $1\%$ of the flows are elephants, considering also that its size is $F_e=100$~MB, due to the fact that distributed file systems usually break long files into $100$-MB size chunks. As we have considered $1$~Gbps paths, for elephant flows we have $\mu^{-1}_{elephant}=F_e/10^9=0.8s$. For mouse flows, and following~\cite{wilson}, we have considered that their flow size is uniformly distributed between $F_m=[2KB,50KB]$, so $\mu^{-1}_{mice}=F_m/10^9=[16,400]\mu s$. From~\cite{wilson}, we have modeled the flow arrival process as a Poisson process, varying the mean arrival rate to get a certain $\rho$. For this scenario, results are shown in Fig.~\ref{fig:simul6}, considering $C=20~\forall i$, $i\in [1,6]$. First of all, it is important to note that results have been obtained for a wide range of scenarios. We depict the loss probability ($LP$) to show such variety of traffic load. Moreover, we can conclude that in this scenario, load is evenly distributed, as the utilization for all paths is very similar. Moreover, we also show the Jain's fairnex index~\cite{jain} ($FI$) which can be defined in our case by 
\[
FI=\frac{(\sum_{i=1}^N{u_i})^2}{6\sum_{i=1}^N{u_i^2}}.
\]
From Jain's fairness index we can conclude that load is evenly distributed with a very high precision, since $FI=1$ represents an optimal load distribution.

\section{Related Work}
\label{related}

Regarding layer 2 switching, the traditional \textbf{Spanning Tree Protocol (STP/RSTP)}~\cite{MACBridges} severely limits the network size and performance by blocking redundant links to prevent loops, thus limiting infrastructure utilization and increasing latency. Successor standards like \textbf{SPB}~\cite{Allan12} or \textbf{TRILL RBridges}~\cite{RBridges,Perlman11} move towards layer 3, e.g. adding link-state control protocols or additional header fields, thus leaving some of the layer 2 benefits behind, such as simplicity or plug-and-play installation. More specifically, SPB is more oriented to interconnection of provider networks than to data center and campus networks, while TRILL RBridges~\cite{RBridges,Perlman11}, standardized by IETF, use a special encapsulation header that is modified at every RBridge hop and is neither compatible with existing switch chipsets nor IEEE OAM nor 802.1aq standards. Moreover, these protocols distribute load statically by hashing the different flows, irrespective of their load status~\cite{AlFares10}.

\textbf{PAST}~\cite{past12} builds a spanning tree per destination host and outperforms standard protocols, but it is based on pre-calculating the routes, lack the dynamicity of All-Path, which considers the path load.
\textbf{ROME}~\cite{rome16}, taking the concepts from \textbf{Greedy Routing}~\cite{greedy16}, presents an architecture and a protocol backwards-compatible with Ethernet, highly scalable and good performance. Nevertheless, it still requires pre-computing the paths via periodical exchange of information among the switches.
\textbf{AXE}~\cite{axe15,axe16} was proposed to recover this simple flood-and-learn mechanism from Ethernet switches. However, it requires the modification of the standard frame by including a hop count and a nonce field. 
\textbf{SynRace}~\cite{synrace15} profits from TCP's congestion control dynamics to select the least-congested paths, by sending probe packets in a similar way to the All-Path protocols. Although the accuracy of SynRace is higher, its produced overhead (table entries, computation of probe packets, etc.) is also much larger.
\textbf{First-Come First-Serve (FCFS)}~\cite{fcfs12} is so far the closest approach to the All-Path family, but its routing tables are more complex (it needs to save the Frame Check Sequence field for every unlearnt packet) and their entries have no refresh option, expiring after a while even if the associated paths are still valid. Moreover, FCFS creates paths analogously to ARP-Path, lacking alternative options similar to Flow-Path or Bridge-Path.

Finally, other proposals might profit from SDN features to create optimal paths by measuring the load, for example. However, these centrallized approaches lack other benefits, such as scalability. In the case of the All-Path family, the ARP-Path protocol was implemented as a hybrid switch taking the best of each world~\cite{aoss17}, proving that this family of protocols can also be combined with SDN if required.

\section{Conclusions}
\label{conclusions}

The All-Path protocols are a family of Ethernet switching protocols that create routes following the lowest latency paths and, at the same time, distributing traffic evenly. These protocols are suitable for campus and data center networks. The family comprises several variants with different advantages in terms of granularity for load balancing and scalability, being ARP-Path, Flow-Path and Bridge-Path.
ARP-Path, the first protocol of the family, creates a path per final host by exploring the whole network. On the one hand, Flow-Path offers even better load balance capabilities and per flow path independence at the cost of bigger table sizes when there are multiple equal cost shortest paths. On the other hand, Bridge-Path provides increased scalability with coarser path granularity specially when the ratio of edge bridges to total number of bridges is high and the number of attached hosts is high.
Finally, we have evaluated the load balancing capabilities of the All-Path family by means of an analytical and simulation models, concluding that the All-Path family protocols are able to use all possible paths evenly.

\section*{Acknowledgment}
This work has been supported by Comunidad de Madrid through project MEDIANET (S-2009/TIC-1468) and project TIGRE5-CM (S2013/ICE-2919).

\ifCLASSOPTIONcaptionsoff
  \newpage
\fi

\bibliographystyle{IEEEtran}
\bibliography{AllPathFamily}   

\end{document}